\documentclass[reprint,superscriptaddress,preprintnumbers,aps,prl]{revtex4-2}
\usepackage{amsmath} 
\usepackage{amssymb}
\usepackage{graphicx,epsfig}
\usepackage[mathscr]{eucal}
\usepackage{color}
\usepackage{longtable}
\usepackage{natbib}
\usepackage[english]{babel}
\usepackage{tabularx}
\usepackage{setspace}
\usepackage{xspace}
\usepackage{multirow}
\renewcommand{\thefigure}{\arabic{figure}}

\newcommand\as{\alpha_{\mathrm{S}}}

\def\to{\rightarrow}

\usepackage{scalefnt,pstricks}

\def\rcut{r_{\rm cut}}

\newcommand{\Qgraf}{{\sc Qgraf}\xspace}
\newcommand{\form}{{\sc Form}\xspace}
\newcommand{\Mathematica}{{\sc Mathematica}\xspace}

\usepackage{array}
\newcolumntype{L}[1]{>{\raggedright\let\newline\\\arraybackslash\hspace{0pt}}m{#1}}
\newcolumntype{C}[1]{>{\centering\let\newline\\\arraybackslash\hspace{0pt}}m{#1}}
\newcolumntype{R}[1]{>{\raggedleft\let\newline\\\arraybackslash\hspace{0pt}}m{#1}}

\usepackage[colorlinks=true,allcolors={blue!70!black}]{hyperref}

\begin{document} 
\preprint{ZU-TH 28/21, TIF-UNIMI-2021-10}
\title{Mixed strong--electroweak corrections to the Drell--Yan process}
\author{Roberto Bonciani}
\email{roberto.bonciani@roma1.infn.it}
\affiliation{Dipartimento di Fisica, Universit\`a di Roma “La Sapienza” and INFN, Sezione di Roma, I-00185 Roma, Italy}
\author{Luca Buonocore}
\email{lbuono@physik.uzh.ch}
\affiliation{Physik Institut, Universit\"at Z\"urich, CH-8057 Z\"urich, Switzerland}
\author{Massimiliano Grazzini}
\email{grazzini@physik.uzh.ch}
\affiliation{Physik Institut, Universit\"at Z\"urich, CH-8057 Z\"urich, Switzerland}
\author{Stefan~Kallweit}
\email{stefan.kallweit@cern.ch}
\affiliation{Dipartimento di Fisica, Universit\`{a} degli Studi di Milano-Bicocca and INFN, Sezione di Milano-Bicocca, I-20126, Milan, Italy}
\author{Narayan Rana}
\email{narayan.rana@mi.infn.it}
\affiliation{Dipartimento di Fisica “Aldo Pontremoli”, University of Milano and INFN, Sezione di Milano, I-20133 Milano, Italy}
\author{Francesco Tramontano}
\email{tramonta@na.infn.it}
\affiliation{Dipartimento di Fisica, Universit\`a di Napoli Federico II and INFN, Sezione di Napoli, I-80126 Napoli, Italy}
\author{Alessandro Vicini}
\email{Alessandro.Vicini@mi.infn.it}
\affiliation{Dipartimento di Fisica “Aldo Pontremoli”, University of Milano and INFN, Sezione di Milano, I-20133 Milano, Italy}
\begin{abstract}
We report on the first complete computation of the mixed QCD--electroweak~(EW) corrections to the neutral-current Drell--Yan process.
Superseding previously applied approximations, our calculation provides the first result at this order that is valid in the entire range of dilepton invariant masses.
The two-loop virtual contribution is computed by using semi-analytical techniques, overcoming the technical problems in the evaluation of the relevant master integrals.
The cancellation of soft and collinear singularities is achieved by a formulation of the $q_T$ subtraction formalism
valid in presence of charged massive particles in the final state.
We present numerical results for the fiducial cross section and selected kinematical distributions.
At large values of the lepton $p_T$ the mixed QCD--EW
corrections are negative and increase in size, to about $-15\%$ with respect to the next-to-leading-order QCD result at \mbox{$p_T=500$\,GeV}.
Up to dilepton invariant masses of 1 TeV the computed corrections amount to about $-1.5\%$ with respect to the next-to-leading-order QCD result.
\end{abstract}
\maketitle

\section{Introduction}

When the Large Hadron Collider at CERN started data taking in 2009, it was expected to give answers to questions like the origin of electroweak~(EW) symmetry breaking or the existence of supersymmetry.
After successful physics runs at 7, 8 and 13 TeV and the discovery of the Higgs boson~\cite{Aad:2012tfa,Chatrchyan:2012ufa}, there is still no clear evidence of physics beyond the Standard Model.
Although a huge amount of data will be accumulated in the high-luminosity phase and exciting discoveries are still well possible,
it is by now clear that an alternative path to uncover possible new physics is the search for small deviations from the predictions of the Standard Model,
and that precision is the key for this path.

The Drell--Yan~(DY) process~\cite{Drell:1970wh} is the perfect example of a precision benchmark process at the LHC.
It corresponds to the inclusive production of a lepton pair through an off-shell vector boson.
It provides large production rates and clean experimental signatures, given the presence of at least one lepton with large transverse momentum in the final state.
Historically, it offered the first application of parton model ideas beyond deep inelastic scattering
and led to the discovery of the $W$ and $Z$ bosons~\cite{Arnison:1983rp,Banner:1983jy,Arnison:1983mk,Bagnaia:1983zx}.
At present, the DY process provides valuable information about parton distribution functions, allows for the precise determination of several Standard Model parameters~\cite{Group:2012gb,Aaboud:2017svj,Aaltonen:2018dxj,ATLAS:2018gqq},
and severely constrains many new-physics scenarios.

The DY process was one of the first hadronic reactions for which radiative corrections in the strong and EW couplings $\as$ and $\alpha$ were computed.
The classic calculations of the next-to-leading-order~(NLO)~\cite{Altarelli:1979ub}
and next-to-next-to-leading-order~(NNLO)~\cite{Hamberg:1990np,Harlander:2002wh} corrections to the total cross section in Quantum Chromodynamics~(QCD) were followed by
(fully) differential NNLO computations including the leptonic decay of the vector boson~\cite{Anastasiou:2003yy,Anastasiou:2003ds,Melnikov:2006kv,Catani:2009sm,Catani:2010en}. 
The complete EW corrections for $W$ production have been computed in Refs.~\cite{Dittmaier:2001ay,Baur:2004ig,Zykunov:2006yb,Arbuzov:2005dd,CarloniCalame:2006zq},
and for $Z$ production in Refs.~\cite{Baur:2001ze,Zykunov:2005tc,CarloniCalame:2007cd,Arbuzov:2007db,Dittmaier:2009cr}.
Very recently, the next-to-next-to-next-to-leading-order~(N$^3$LO) QCD radiative calculations of the inclusive production of
a virtual photon~\cite{Duhr:2020seh} and of a $W$ boson~\cite{Duhr:2020sdp} have been completed, and first estimates of fiducial cross sections for the neutral-current DY process at the same order have appeared~\cite{Camarda:2021ict}.

Since the high-precision determination of EW parameters requires control over the kinematical distributions at very high accuracy,
the attention of the theory community has recently turned to the mixed QCD--EW corrections.
The knowledge of these corrections would indeed allow us
to improve over the approximations offered by shower Monte Carlo programs~\cite{Barze:2013fru,Frederix:2018nkq},
which include only partial subsets of factorisable mixed QCD--EW corrections, and to reduce the remaining theoretical uncertainties.

The mixed QCD--QED corrections to the inclusive production of an on-shell $Z$ boson were obtained
in Ref.~\cite{deFlorian:2018wcj} through an abelianisation procedure from the NNLO QCD results~\cite{Hamberg:1990np,Harlander:2002wh}.
This calculation was extended to the fully differential level for off-shell $Z$ boson production and decay into a pair of neutrinos (i.e.\ without final-state radiation) in Ref.~\cite{Cieri:2020ikq}.
A similar calculation was carried out in Ref.~\cite{Delto:2019ewv} in an on-shell approximation for the $Z$ boson, but including the factorised NLO QCD corrections to $Z$ production and the NLO QED corrections to the leptonic $Z$ decay.
Complete ${\cal O}(\as\alpha)$ computations for the production of on-shell $Z$ and $W$ bosons have been presented
in Refs.~\cite{Bonciani:2016wya,Bonciani:2019nuy,Bonciani:2020tvf,Buccioni:2020cfi,Behring:2020cqi}.
Beyond the on-shell approximation, the most relevant results have been obtained in the {\it pole} approximation~\cite{Denner:2019vbn}.
This approximation is based on a systematic expansion of the cross section around the $W$ or $Z$ resonance, in order to split the radiative corrections
into well-defined, gauge-invariant contributions. 
Such method has been used in Refs.~\cite{Dittmaier:2014qza,Dittmaier:2015rxo} to evaluate what is expected to be the dominant part of the mixed QCD--EW corrections in the resonance region.

Given the relevance of mixed QCD--EW corrections for precision studies of DY production and for an accurate measurement of the $W$ mass~\cite{CarloniCalame:2016ouw,Behring:2021adr},
it is important to go beyond this approximation.
New-physics effects, in particular, could manifest themselves in the tails of kinematical distributions, where the pole approximation is not expected to work.
A first step in this direction has been carried out in Ref.~\cite{Dittmaier:2020vra}, where complete
results for the ${\cal O}(n_F\as \alpha)$ contributions to the DY cross section were presented.
Very recently, some of us have presented a computation~\cite{Buonocore:2021rxx} of the mixed QCD--EW corrections to the charged-current process
$pp\to \ell \nu_\ell+X$, where all contributions are evaluated exactly except for the finite part of the two-loop amplitude, which was evaluated in the pole approximation.

One of the bottlenecks for a complete ${\cal O}(\as\alpha)$ calculation is indeed the corresponding two-loop virtual amplitude. 
The evaluation of the $2\to 2$ two-loop Feynman diagrams with internal masses is at the frontier of current computational techniques.
Progress on the evaluation of the corresponding two-loop master integrals has been reported in Refs.~\cite{Bonciani:2016ypc,Heller:2019gkq,Hasan:2020vwn,Bonciani:2008az,Bonciani:2009nb,Mastrolia:2017pfy}. Very recently,
the computation of the two-loop helicity amplitudes for neutral-current massless lepton pair production was discussed in Ref.~\cite{Heller:2020owb}.
In this Letter we report on an independent calculation of the two-loop amplitude
and on its combination with the remaining perturbative contributions,
to obtain the first complete computation
of the mixed QCD--EW corrections for the neutral-current DY process.

\section{The calculation}

We consider the inclusive production of a charged-lepton pair in proton collisions,
\begin{equation}
  pp\to \ell^+\ell^-+X\, .
  \label{eq:proc}
  \end{equation}
The theoretical predictions for this process can be obtained as a convolution of the parton distribution functions for the incoming protons with the hard scattering partonic cross section.
When QCD and EW radiative corrections are considered, the initial partons include (anti-)quarks, gluons and photons.

The differential cross section for the process in Eq.~(\ref{eq:proc}) can be written as
\begin{equation}
  \label{eq:exp}
  d{\sigma}=\sum_{m,n=0}^\infty d{\sigma}^{(m,n)}\, ,
\end{equation}
where \mbox{$d{\sigma}^{(0,0)}\equiv d{\sigma}_{\rm LO}$} is the Born level contribution and $d{\sigma}^{(m,n)}$ the ${\cal O}(\as^m\alpha^n)$ correction.
The mixed QCD--EW corrections correspond to the term \mbox{$m=n=1$} in this expansion and include double-real, real--virtual and purely virtual contributions.
The corresponding tree-level and one-loop scattering amplitudes are computed with {\sc Openloops}~\cite{Cascioli:2011va, Buccioni:2017yxi, Buccioni:2019sur} and {\sc Recola}~\cite{Actis:2016mpe,Denner:2017wsf}, finding complete agreement.
The two-loop amplitude is computed using the following method.
The Feynman diagrams are generated with \Qgraf~\cite{Nogueira:1991ex}. Using an in-house \form~\cite{Vermaseren:2000nd} program we compute the unpolarized interference with the tree-level diagrams.
The computation is done consistently in $d$ space-time dimensions,
using a naive anticommuting $\gamma_5$~\cite{Kreimer:1989ke} and the scheme proposed in Ref.~\cite{Larin:1993tq}.
The interference is therefore expressed in terms of dimensionally regularized scalar integrals that are reduced to the master integrals~(MIs)
using integration-by-parts~\cite{Tkachov:1981wb,Chetyrkin:1981qh} and Lorentz-invariance~\cite{Gehrmann:1999as} identities, as implemented in the computer codes
{\sc Kira}~\cite{Maierhoefer:2017hyi}, {\sc LiteRed}~\cite{Lee:2012cn} and {\sc Reduze~2}~\cite{Studerus:2009ye,vonManteuffel:2012np}.
The resulting set of MIs are available in the literature~\cite{Bonciani:2008az,Bonciani:2009nb,Mastrolia:2017pfy,Bonciani:2016ypc,Heller:2019gkq,Hasan:2020vwn}. In particular, for the MIs with massive-boson exchange we refer to the implementation given in Ref.~\cite{Bonciani:2016ypc}, where they are expressed in terms of generalized polylogarithms and Chen-iterated integrals~\cite{Goncharov:polylog,Goncharov2001,Remiddi:1999ew,Chen:1977oja}. The numerical evaluation of the Chen-iterated structures is very complicated and not a viable solution for a practical implementation. Therefore, for the evaluation of the most complicated MIs (five two-loop box-type MIs with two massive lines) we employ
the semi-analytical method of Ref.~\cite{Francesco:2019yqt}, implemented in the 
\Mathematica-based program {\sc DiffExp}~\cite{Hidding:2020ytt}.
Numerical checks have been performed with
{\sc Fiesta}~\cite{Smirnov:2015mct} and py{\sc SecDec}~\cite{Borowka:2017idc}.

The computation of the amplitude is organised by breaking it into different gauge-independent ultraviolet-renormalised subsets of diagrams, defined by the different possible combinations of electric and weak charges.
In the evaluation of the amplitude we keep the lepton mass wherever needed to regularize the final-state collinear singularities
\footnote{We have explicitly checked the cancellation of the collinear divergences arising in the box diagrams with a photon exchanged between the quark and lepton lines~\cite{Frenkel:1976bj}.}.
The two-loop virtual amplitude is computed in the background-field gauge~\cite{Denner:1994xt},
which restores the validity of QED-like Ward identities in the full SM.
The evaluation of the relevant two-loop counterterms~\cite{Sirlin:1980nh,Degrassi:2003rw}
is given in terms of two-loop self-energy diagrams~\cite{Kniehl:1989yc,Djouadi:1993ss,Dittmaier:2020vra}. 

Even when all the amplitudes have been computed, the completion of the calculation remains a formidable task. Indeed, double-real, real--virtual and purely virtual contributions are separately infrared divergent,
and a method to handle and cancel infrared singularities has to be worked out.
In this work we use a formulation of the $q_T$ subtraction formalism~\cite{Catani:2007vq} derived from the NNLO QCD computation of heavy-quark production~\cite{Catani:2019iny,Catani:2019hip,Catani:2020kkl} through an appropriate abelianisation procedure~\cite{deFlorian:2018wcj,Buonocore:2019puv}. The same method has been recently applied to the charged-current DY process~\cite{Buonocore:2021rxx}.
According to the $q_T$ subtraction formalism~\cite{Catani:2007vq} $d{\sigma}^{(m,n)}$ can be evaluated as
\begin{equation}
\label{eq:master}
d{\sigma}^{(m,n)}={\cal H}^{(m,n)}\otimes d{\sigma}_{\rm LO}+\left[d\sigma_{\rm R}^{(m,n)}-d\sigma_{\rm CT}^{(m,n)}\right]\, .
\end{equation}
The first term in Eq.~(\ref{eq:master}) is obtained through a convolution (denoted by the symbol $\otimes$)
of the perturbatively computable function ${\cal H}^{(m,n)}$ and the LO cross section $d{\sigma}_{\rm LO}$,
with respect to the longitudinal-momentum fractions of the colliding partons.
The second term is the {\it real} contribution $d\sigma_{\rm R}^{(m,n)}$, where the charged leptons
are accompanied by additional QCD and/or QED radiation that produces a recoil with finite transverse momentum $q_T$.
For \mbox{$m+n=2$} such contribution can be evaluated by using the dipole subtraction formalism~\cite{Catani:1996jh,Catani:1996vz,Catani:2002hc,Kallweit:2017khh,Dittmaier:1999mb,Dittmaier:2008md,Gehrmann:2010ry,Schonherr:2017qcj}.
In the limit $q_T\to 0$ the real contribution
$d\sigma_{\rm R}^{(m,n)}$ is divergent, since the recoiling radiation becomes soft and/or collinear to the initial-state partons.
Such divergence is cancelled by the counterterm $d\sigma_{\rm CT}^{(m,n)}$, which eventually makes the cross section in Eq.~(\ref{eq:master}) finite.

The required phase space generation and integration is carried out within the {\sc Matrix} framework~\cite{Grazzini:2017mhc}.
The core of {\sc Matrix} is the Monte Carlo program {\sc Munich}~\footnote{{\sc Munich}, which is the abbreviation of “MUlti-chaNnel Integrator at Swiss~(CH) precision”, is an automated parton-level NLO\
generator by S. Kallweit.}, which contains a fully automated implementation of the dipole subtraction method for massless and massive partons
at NLO QCD~\cite{Catani:1996jh,Catani:1996vz,Catani:2002hc} and NLO EW~\cite{Kallweit:2017khh,Dittmaier:1999mb,Dittmaier:2008md,Gehrmann:2010ry,Schonherr:2017qcj}.
The $q_T$ subtraction method has been applied to several NNLO QCD computations for the production of colourless final-state systems (see Ref.~\cite{Grazzini:2017mhc} and references therein), and to heavy-quark production~\cite{Catani:2019iny,Catani:2019hip,Catani:2020kkl}, which correspond to the case \mbox{$m=2$}, \mbox{$n=0$}.
The method has also been applied in Ref.~\cite{Buonocore:2019puv} to study NLO EW corrections to the DY process, which represents the case \mbox{$m=0$}, \mbox{$n=1$}.
Very recently, some of us have applied the method to the computation of mixed QCD--EW corrections to the charged-current DY process~\cite{Buonocore:2021rxx}.
The structure of the coefficients ${\cal H}^{(1,1)}$ and $d\sigma_{\rm CT}^{(1,1)}$ can be derived from those controlling
the NNLO QCD computation of heavy-quark production.
The initial-state soft/collinear and purely collinear contributions were already presented in Ref.~\cite{Cieri:2020ikq}.
The fact that the final state is colour neutral implies that final-state radiation is of pure QED origin.
Therefore, the purely soft contributions
have a simpler structure than the corresponding contributions entering the
NNLO QCD computation of Refs.~\cite{Catani:2019iny,Catani:2019hip,Catani:2020kkl}.
The final result for the infrared-subtracted two-loop contribution, which enters the coefficient ${\cal H}^{(1,1)}$,
is evaluated numerically on a 
two-dimensional grid by using the tools {\sc HarmonicSums}~\cite{Ablinger:2010kw, Ablinger:2014rba}, {\sc Ginac}~\cite{Vollinga:2004sn} and {\sc PolyLogTools}~\cite{Duhr:2019tlz}.

\section{Results}

We consider the process \mbox{$pp\to \mu^+\mu^-+X$} at the centre-of-mass energy \mbox{$\sqrt{s}=14$\,TeV}. 
As for the EW couplings, we follow the setup of Ref.~\cite{Dittmaier:2015rxo}.
In particular, we use the $G_\mu$ scheme with \mbox{$G_F=1.1663787\times 10^{-5}$\,GeV$^{-2}$} and set the \textit{on-shell} values of masses and widths to \mbox{$m_{W,{\rm OS}}=80.385$\,GeV}, \mbox{$m_{Z, {\rm OS}}=91.1876$\,GeV}, \mbox{$\Gamma_{W, {\rm OS}}=2.085$\,GeV}, \mbox{$\Gamma_{Z, {\rm OS}}=2.4952$\,GeV}. Those values are translated to the corresponding \textit{pole} values \mbox{$m_{V}=m_{V,{\rm OS}}/\sqrt{1+\Gamma^2_{V,{\rm OS}}/m^2_{V{,\rm OS}}}$} and \mbox{$\Gamma_{V}=\Gamma_{V,{\rm OS}}/\sqrt{1+\Gamma^2_{V,{\rm OS}}/m_{V,{\rm OS}}^2}$}, \mbox{$V=W,Z$}, from which  \mbox{$\alpha=\sqrt{2}\,G_F m_{W}^2(1-m_{W}^2/m_{Z}^2)/\pi$} is derived,
and we use the complex-mass scheme~\cite{Denner:2005fg} throughout~\footnote{For a technical limitation of the semi-analytical approach, the evaluation of the box-type Feynman diagrams with two internal massive lines has been carried out with real masses of the gauge bosons in the Feynman integrals.}.
The muon mass is fixed to \mbox{$m_\mu=105.658369$\,MeV}, and the pole masses of the top quark and the Higgs boson to \mbox{$m_t=173.07$\,GeV} and \mbox{$m_H=125.9$\,GeV}, respectively.
The CKM matrix is taken to be diagonal.
We work with \mbox{$n_f=5$} massless quark flavours and retain the exact top-mass dependence in all virtual and real--virtual
amplitudes associated to bottom-induced processes, except for the two-loop virtual
corrections,  where we neglect top-mass effects.
Given the smallness of the bottom-quark density, we estimate the corresponding error to be at the percent level of the computed correction.
We use the \texttt{NNPDF31$\_$nnlo$\_$as$\_$0118$\_$luxqed} set of parton distributions~\cite{Bertone:2017bme}, which is based on the LUXqed methodology~\cite{Manohar:2016nzj}
for the determination of the photon density. Correspondingly, the QCD coupling $\as$ is evaluated at three-loop order.
The renormalisation and factorisation scales are fixed to \mbox{$\mu_R=\mu_F=m_Z$}.

We use the following selection cuts on the transverse momenta and rapidities of the muons, $p_{T,\mu^\pm}$ and $y_{\mu^\pm}$, and on the invariant mass $m_{\mu\mu}$ of the di-muon pair,
\begin{equation}
  \label{eq:cuts}
p_{T,\mu^\pm}>25\,{\rm GeV}\,,\quad |y_{\mu^\pm}|<2.5\,,\quad m_{\mu\mu}>50\,{\rm GeV}\, .
\end{equation}
We work at the level of {\it bare muons}, i.e., no lepton recombination with close-by photons is carried out.

We start the presentation of our results with the fiducial cross section. In Table~\ref{tab:fid} we report
the contributions $\sigma^{(i,j)}$ to the cross section (see Eq.~(\ref{eq:exp})) in the various partonic channels. The numerical uncertainties are stated in brackets, and for the NNLO corrections
$\sigma^{(2,0)}$ and the mixed QCD--EW contributions $\sigma^{(1,1)}$ they include the systematic uncertainties that will be discussed below.
The contribution from quark--antiquark annihilation is denoted by $q{\bar q}$.
\begin{table}
  \small
\renewcommand{\arraystretch}{1.4}
  \centering
  \begin{tabular}{|c|c|c|c|c|c|}
    \hline
    $\sigma$ [pb] & $\sigma_{\rm LO}$  & $\sigma^{(1,0)}$  & $\sigma^{(0,1)}$  & $\sigma^{(2,0)}$ & $\sigma^{(1,1)}$ \\
    \hline
    \hline
    $q{\bar q}$ & $809.56(1)$ & $191.85(1)$ & $-33.76(1)$ & $49.9(7)$ & $-4.8(3)$\\
    \hline
    $qg$ & --- & $-158.08(2)$ & --- & $-74.8(5)$ & $8.6(1)$ \\
    \hline
    $q(g)\gamma$ & --- & --- & $-0.839(2)$ & --- & $0.084(3)$\\
    \hline
    $q({\bar q})q^\prime$ & --- & --- & --- & $6.3(1)$ & $0.19(0)$\\
    \hline
    $g g$ & --- & --- & --- & $18.1(2)$ & --- \\
     \hline
    $\gamma \gamma$ & $1.42(0)$ & --- & $-0.0117(4)$ & --- & --- \\
    \hline
    \hline
    tot & $810.98(1)$ & $33.77(2)$ & $-34.61(1)$ & $-0.5(9)$ &  $4.0(3)$ \\
    \hline
  \end{tabular}
  \caption{\label{tab:fid}
    The different perturbative contributions to the fiducial cross section (see Eq.~(\ref{eq:exp})). The breakdown into the various partonic channels is also shown (see text).}
\renewcommand{\arraystretch}{1.0}
  \end{table}
The contributions from the channels
\mbox{$qg+{\bar q}g$} and \mbox{$q\gamma+{\bar q}\gamma+g\gamma$}
are labelled by $qg$ and $q(g)\gamma$, respectively.
The contribution from all the remaining quark--quark channels
$qq',\, {\bar q}{\bar q}'$ (including both \mbox{$q=q'$} and \mbox{$q\neq q'$}) and $q{\bar q}'$ (with \mbox{$q\neq q'$})
is labelled by $q({\bar q})q^\prime$.
Finally, the contributions from the gluon--gluon and photon--photon  channels are denoted by $gg$ and $\gamma\gamma$, respectively.
We see that radiative corrections are subject to large cancellations between the various partonic channels. The NLO QCD corrections amount to $+4.2\%$ with respect to the LO result, while the NLO EW corrections contribute $-4.3\%$. Also the NNLO QCD corrections are subject to large cancellations, and give an essentially vanishing contribution within the numerical uncertainties. The newly computed QCD--EW corrections amount to $+0.5\%$ with respect to the LO result.

\begin{figure}[t]
\begin{center}
\includegraphics[width=\columnwidth]{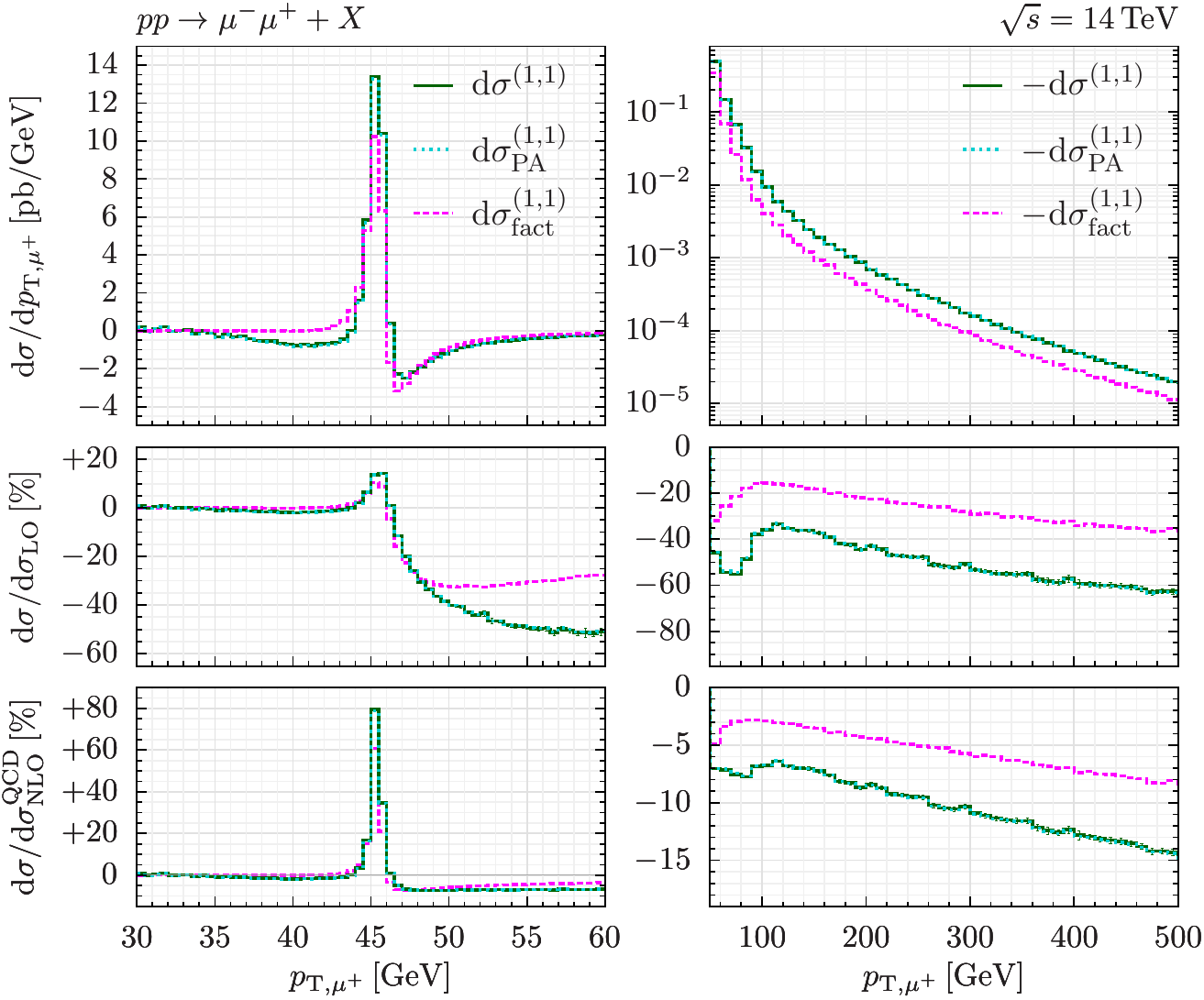}\\[2ex]
\end{center}
\vspace{-2ex}
\caption{\label{fig:pt}
Complete ${\cal O}(\as\alpha)$ correction to the differential cross section $d\sigma^{(1,1)}$ in the anti-muon $p_T$ compared to the corresponding result in the pole approximation and to the
factorised approximation $d\sigma^{(1,1)}_{\rm fact}$.
The top panels show the absolute predictions, while the central (bottom) panels display the ${\cal O}(\as\alpha)$ correction normalized to the LO (NLO QCD) result.
For the full result the ratios also display our estimate of the numerical uncertainties, obtained as described in the text.
}
\end{figure}
In Fig.~\ref{fig:pt} we present our result for the ${\cal O}(\as\alpha)$ correction as a function of the anti-muon $p_T$.
The left panels depict the region around the $Z$ peak, and the right panels the high-$p_T$ region.
In the main panels we show the absolute correction $d\sigma^{(1,1)}/dp_T$,
while the central (bottom) panels display the correction normalised to the LO (NLO QCD) result.
Our results for the complete ${\cal O}(\as\alpha)$ correction are compared with those obtained in two approximations.
The first approximation consists in computing the finite part of the two-loop virtual amplitude in the pole approximation, suitably reweighted
with the exact squared Born amplitude.
This approach precisely follows that adopted for the charged-current DY process in Ref.~\cite{Buonocore:2021rxx} (see Eq.~(14) therein for the precise definition).
The pole approximation, which includes factorisable and non-factorisable~\cite{Dittmaier:2014qza} contributions, requires the QCD--EW on-shell form factor of the $Z$ boson~\cite{Bonciani:2020tvf}.
The second approximation is based on a fully factorised approach for QCD and EW corrections, where we exclude photon-induced processes throughout (see Ref.~\cite{Dittmaier:2015rxo,Buonocore:2021rxx} for a detailed description).
We see that the result obtained in the pole approximation is in perfect agreement with the exact result. This is due to the small contribution of the two-loop virtual to the computed correction,
as observed also in the case of $W$ production~\cite{Buonocore:2021rxx}.
Our result for the ${\cal O}(\as\alpha)$ correction in the region of the peak is reproduced relatively well by the factorised approximation.
Beyond the Jacobian peak, this approximation tends to overshoot the complete result,
which is consistent with what was observed in Refs.~\cite{Dittmaier:2015rxo,Buonocore:2021rxx}. As $p_T$ increases, the (negative) impact of the mixed QCD--EW corrections increases, and at \mbox{$p_T=500$\,GeV} it reaches about $-60\%$ with respect to the LO prediction and $-15\%$ with respect to the NLO QCD result.
The factorised approximation describes the qualitative behaviour of the complete correction reasonably well, also in the tail of the distribution, but it overshoots the full result as $p_T$ increases.

\begin{figure}[t]
\begin{center}
\includegraphics[width=\columnwidth]{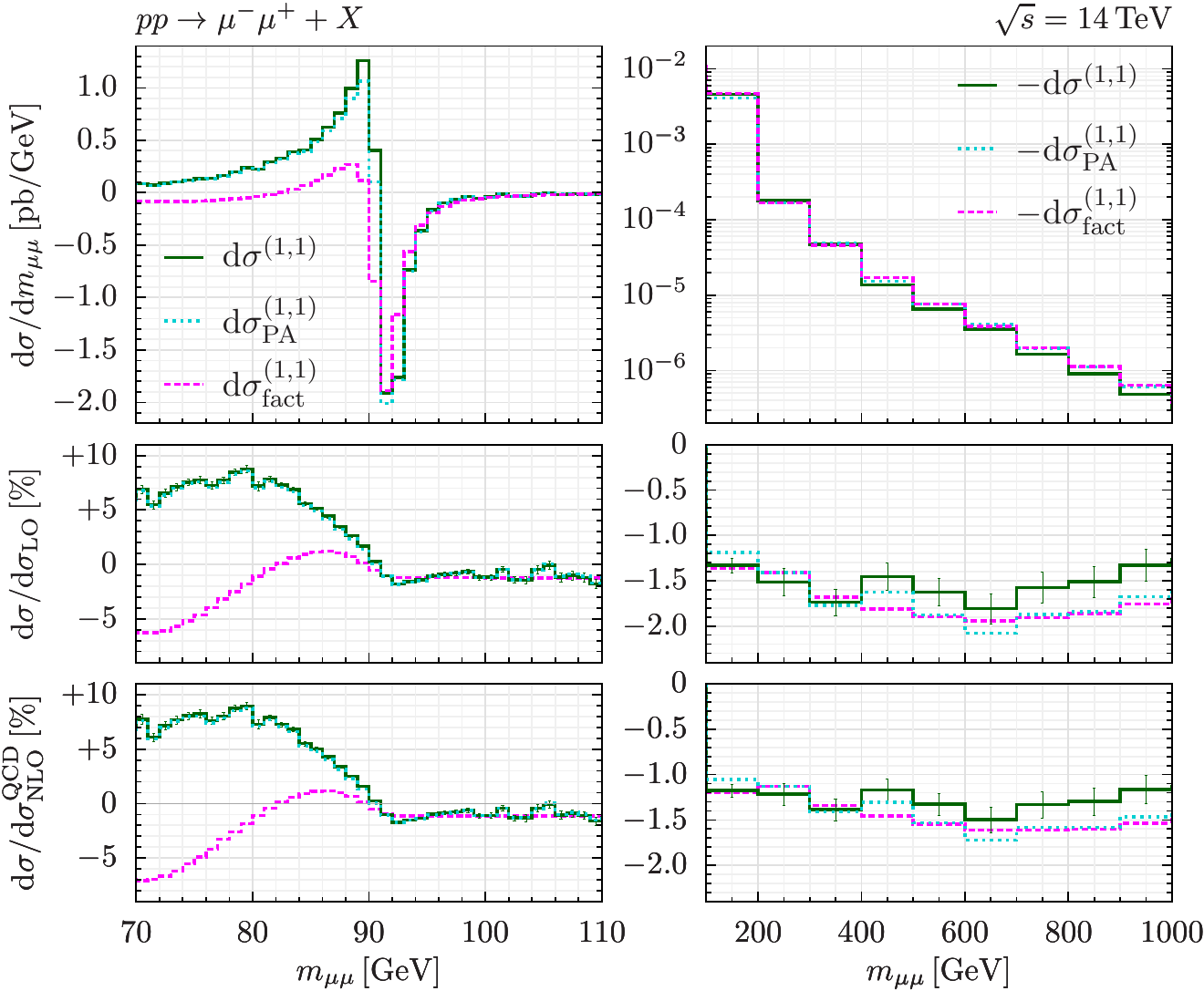}\\[2ex]
\end{center}
\vspace{-2ex}
\caption{\label{fig:mll}
As Fig.~\ref{fig:pt}, but for the di-muon invariant mass.
}
\end{figure}
In Fig.~\ref{fig:mll} we show our result for the ${\cal O}(\as\alpha)$ correction as a function of the di-muon invariant mass $m_{\mu\mu}$.
The left panels depict the region around the $Z$ peak, and the right panels the high-$m_{\mu\mu}$ region.
When comparing the factorised approximation with the exact result,
we notice that it fails to describe the radiative correction below the $Z$ resonance, as already pointed out in Ref.~\cite{Dittmaier:2015rxo}.
In contrast, the pole approximation is a very good approximation
of the complete correction, with some small differences that can be appreciated right around the peak.
In the high-$m_{\mu\mu}$ region the correction is uniformly of the order of $-1.5\%$ with respect to the NLO QCD result. Here the trend of the negative correction
is captured by both approximations,
which, however, both undershoot the exact result by about $30\%$, highlighting the relevance of the exact two-loop contribution for this observable.

The numerical evaluation of the above results for $d\sigma^{(1,1)}$ (and $d\sigma^{(2,0)}$) requires the introduction of a
technical cut-off $\rcut$ on the dimensionless variable $q_T/m_{\mu\mu}$ in the square bracket of Eq.~(\ref{eq:master}).
We follow the procedure in {\sc Matrix}~\cite{Grazzini:2017mhc} to simultaneously calculate for several values of $\rcut$
and to perform a numerical extrapolation \mbox{$\rcut\to 0$}, but apply it on a bin-wise level.
Quadratic least $\chi^2$ fits in the range $[0.01\%, r_{\rm max}]$ with $r_{\rm max} \in [0.25\%, 0.5\%]$ are used to determine best predictions and extrapolation error estimates.
In the case of the anti-muon $p_T$ distribution, the final uncertainties of the computed correction, combining statistical and systematic errors, range from the percent level in the peak region to ${\cal O}(3\%)$ in the tail. In the case of the di-muon invariant mass, the final uncertainties are larger, and range from the few percent level in the peak region to ${\cal O}(10\%)$ at high $m_{\mu\mu}$ values.

\section{Summary}

In this Letter we have presented the first complete computation of the
mixed QCD--EW corrections to neutral-current DY lepton pair production at the LHC.
All the real and virtual contributions
due to initial- and final-state radiation
are included exactly,
thereby allowing us to investigate the impact of the computed corrections in the entire region of dilepton invariant masses.
The evaluation of the two-loop virtual amplitude has been achieved by using semi-analytical techniques.
To cancel soft and collinear singularities, we have used a
formulation of the $q_T$ subtraction formalism derived from the NNLO QCD calculation for heavy-quark production through an appropriate abelianisation procedure.
Our computation is fully differential in the momenta of the charged leptons and the associated QED and QCD radiation.
Therefore, it can be used to compute arbitrary infrared-safe observables, and, in particular, we can also deal with {\it dressed leptons}, i.e.\ leptons recombined with close-by photons.
More detailed results of our calculation will be presented elsewhere.

\vskip 1.5cm
\noindent {\bf Acknowledgements}

\noindent

We would like to express our gratitude to Jean-Nicolas Lang and Jonas Lindert for their
continuous support on {\sc Recola} and {\sc OpenLoops}, to Simone Devoto for fruitful discussions and several checks of the two-loop amplitudes, and to Chiara Savoini for numerical checks of the pole approximation.
This work is supported in part by the Swiss National Science Foundation~(SNF) under contract 200020$\_$188464. The work of SK is supported by the ERC Starting Grant 714788 REINVENT. FT acknowledges support from INFN.
AV and NR are supported by the Italian \textit{Ministero dell'Universit\`a e della Ricerca} (Grant No. PRIN2017) and by the European Research Council under the European Unions Horizon 2020 research and innovation Programme (Grant Agreement No. 740006).
RB and NR acknowledge the COST (European Cooperation in Science and Technology) Action CA16201 PARTICLEFACE for partial support. 

\bibliography{biblio}

\end{document}